\DocumentMetadata{
  tagging=on,
  tagging-setup={math/setup=mathml-SE},
  pdfstandard={A-4f,UA-2},
  lang=en-US
}
\documentclass[manuscript]{acmart}

\AtBeginDocument{%
  \providecommand\BibTeX{{%
    Bib\TeX}}}

\usepackage[skins]{tcolorbox}
\usepackage{multirow}
\usepackage{listings}
\usepackage{subcaption}
\usepackage[T1]{fontenc}
\usepackage{fancybox,framed}
\usepackage{xcolor}
\usepackage{graphicx}
\usepackage{longtable}
\usepackage{lscape}
\usepackage{todonotes}
\usepackage{multirow}
\usepackage{color}
\usepackage{colortbl}
\definecolor{Gray}{gray}{0.9}
\usepackage[flushleft]{threeparttable}
\usepackage{listings}
\usepackage{enumitem}
\usepackage{tablefootnote}
\usepackage{float}
\usepackage{tabularx}
\usepackage{xcolor} 
\usepackage{graphicx} 

\usepackage{soul}
\soulregister\cite7
\soulregister\ref7
\def\BibTeX{{\rm B\kern-.05em{\sc i\kern-.025em b}\kern-.08emT\kern-.1667em\lower.7ex\hbox{E}\kern-.125emX}}

\setcopyright{none}
\renewcommand\footnotetextcopyrightpermission[1]{}

\begin{document}

\title[Faster but Not Wiser: GitHub Copilot Decouples Programming Performance from Code Comprehension in Brownfield Tasks]{Faster but Not Wiser: GitHub Copilot Decouples Programming Performance from Code Comprehension in Brownfield Tasks}

\author{Yunhan Qiao}
\orcid{0009-0006-5852-321X}
\email{qiaoy@oregonstate.edu}
\affiliation{%
  \institution{Oregon State University}
  \city{Corvallis}
  \state{Oregon}
  \country{USA}
}

\author{Md Istiak Hossain Shihab}
\orcid{0009-0001-9362-2268}
\email{shihabm@oregonstate.edu}
\affiliation{%
  \institution{Oregon State University}
  \city{Corvallis}
  \state{Oregon}
  \country{USA}
}

\author{Summit Haque}
\orcid{0000-0002-4658-4393}
\email{haquesu@oregonstate.edu}
\affiliation{%
  \institution{Oregon State University}
  \city{Corvallis}
  \state{Oregon}
  \country{USA}
}

\author{Christopher Hundhausen}
\orcid{0000-0001-5128-905X}
\email{chris.hundhausen@oregonstate.edu}
\affiliation{%
  \institution{Oregon State University}
  \city{Corvallis}
  \state{Oregon}
  \country{USA}
}

\renewcommand{\shortauthors}{Qiao et al.}
\newcommand{\AbstractCategory}[1]{%
  \par\addvspace{.5\baselineskip}
  \noindent\textbf{#1}\quad\ignorespaces
}

\begin{abstract}

Teaching Computer Science (CS) students to comprehend and maintain existing codebases is a critical challenge in software engineering education. Although Generative AI (GenAI) assistants such as GitHub Copilot can improve task completion speed and correctness, their relationship with code comprehension remains unclear. We conducted a within-subjects study with 15 CS graduate students who completed feature-implementation tasks in an unfamiliar codebase with and without Copilot. Despite significant performance improvements with Copilot, participants showed no corresponding improvement in overall comprehension ($p=0.59$), and performance gains were not significantly associated with comprehension gains. Exploratory category-level estimates were positive for identifying what and where to modify ($\rho=0.50$) and negative for explaining how the existing code worked and predicting the effects of a change ($\rho=-0.57$); however, neither remained significant after correction for multiple comparisons. Our behavioral analysis showed that participants with higher comprehension engaged more frequently in verification loops, repeatedly inspecting and revising code. They performed write-then-view transitions 4.7 times more often than participants with lower comprehension ($p=0.001$). These findings show that successful task completion does not reliably indicate code comprehension and suggest that how students engage with AI-generated code may matter for their resulting understanding. We argue for programming education that assesses correctness and comprehension separately, teaches students to inspect and explain AI-generated code, and encourages GenAI tools that support active verification.
\end{abstract}




\maketitle

\section{Introduction}
\label{sections-s1-introduction}
Equipping CS students with the ability to comprehend and modify legacy software is a core objective of software engineering education. In industry, developers spend more than half of their time understanding existing source code \cite{xia2017measuring, qiao2026systematic}. However, these so-called \textit{brownfield} programming tasks \cite{baley2010brownfield} remain difficult for students to master, as they require the formation of complex cognitive models \cite{storey2006theories, sillito2006questions} and information foraging strategies \cite{ko2006exploratory}. 

Historically, a student's programming ability has served as a reliable proxy for their understanding of the underlying codebase \cite{peitek2022correlates}. To correctly extend or debug a system, programmers generally need to determine where a change should be made, understand how relevant components interact, and anticipate how the modification will affect program behavior \cite{peitek2022correlates, shaft2006role}. Programming performance and comprehension are not identical, but they have traditionally been closely connected. Consequently, educators often use successful task completion, functional correctness, and passed test cases as indicators that students have developed an adequate understanding of the code they modified.

However, the emergence of Generative AI (GenAI) threatens to break this link. Tools such as GitHub Copilot can generate complete functions, suggest modifications, and translate natural-language requirements into working code. Developers may therefore produce correct solutions without personally performing all of the reasoning previously required to understand and modify an existing system. A student may know what functionality is required and where a change should be inserted while remaining unable to explain how the generated code works or how it interacts with the code base. As a result, GenAI may weaken the traditional relationship between programming performance and code comprehension. 

This possibility poses an important challenge for computing education. Prior studies show that GenAI assistants can improve programming speed and task completion \cite{peng2023impact}, but they also identify concerns about code quality and uneven learning outcomes \cite{he2025speed, prather2024widening}. As programmers increasingly shift from directly producing code to evaluating and integrating AI-generated output, a correct implementation may reflect effective tool use without necessarily demonstrating a corresponding understanding of the code. Artifact correctness may therefore no longer be sufficient evidence of learning. This concern is especially salient in brownfield programming, where developers must understand an existing system well enough to maintain, debug, and extend it. Although prior work has examined GenAI-supported code explanations and students’ use of Copilot in large code bases \cite{nam2024using, yan2024ivie, shihab2025effects}, relatively little is known about how general-purpose GenAI assistants affect code comprehension during authentic brownfield programming tasks \cite{qiao2026systematic, shihab2025effects}. We therefore ask:

\begin{tcolorbox}[
  colback=white,
  colframe=black,
  boxrule=0.6pt,
  left=6pt,
  right=6pt,
  top=4pt,
  bottom=4pt,
  sharp corners,
  width=\columnwidth
]
\emph{Overall RQ:} How does GenAI assistance influence programmers' code comprehension and programming behaviors when completing brownfield programming tasks in unfamiliar code bases?
\end{tcolorbox}

To systematically investigate this overall question, we pose three specific research questions:
\begin{description}
\item[\textbf{RQ1:}] \textit{How does programmers' code comprehension in brownfield programming tasks differ with and without GenAI?}
\item[\textbf{RQ2:}] \textit{Are there tradeoffs between programmers' code comprehension and their task performance with GenAI?}
\item[\textbf{RQ3:}] \textit{How are programming behaviors and activity patterns with GenAI related to higher code comprehension in brownfield programming tasks?}
\end{description}

To address these questions, we conducted a within-subjects study in which participants worked on two brownfield programming tasks with and without GitHub Copilot, completing a comprehension test after each set of tasks. We first replicated prior work~\cite{shihab2025effects}, confirming that Copilot significantly improves task completion speed and test pass rates in our graduate CS population. Our novel contribution lies in examining whether these productivity gains translate into improved code comprehension, a gap in prior work.

Our analysis revealed a \textit{comprehension-performance decoupling}: While participants demonstrated significant productivity gains with Copilot, their comprehension scores did not improve correspondingly; we found no significant correlation between programming productivity improvement and overall code comprehension gains in the Copilot condition. This contrasts with traditional programming scenarios~\cite{peitek2022correlates, shaft2006role}, where comprehension and performance are typically linked. Exploratory category-level correlations showed different estimates for Implementation and Reverse Engineering comprehension, but none remained statistically significant after correction for multiple comparisons. A follow-up behavioral analysis revealed a key insight: how programmers interacted with Copilot showed strong associations with their overall comprehension levels. These findings suggest that the relationship between GenAI assistance, comprehension, and productivity is more nuanced than in traditional programming contexts, with important implications for both programming education and the design of GenAI programming assistants.

\section{Related Work}
\label{sections-s2-related-work}
\subsection{Code Comprehension}

Code comprehension is a critical activity in software development. Previous studies estimate that developers spend 52--70\% of their time understanding existing source code before modifying it
\cite{al2017source, koenemann1991expert, xia2017measuring}.
To construct a mental model of a software system, developers may employ a top-down strategy \cite{brooks1977towards}, a bottom-up strategy \cite{letovsky1987cognitive}, or an integrated strategy that combines both approaches \cite{heinonen2023synthesizing}.

Empirical studies show that comprehension in practice is more dynamic than these models alone suggest. Von Mayrhauser et al. observed professional developers completing software enhancement tasks and found that they repeatedly moved among different parts of a system and different levels of abstraction \cite{von1997program}. They therefore characterized comprehension as a multi-layered, dynamic, and iterative process. Levy and Feitelson similarly found that understanding large-scale software requires developers to gradually construct layered mental models of the system's structure and behavior \cite{levy2019understanding}.

Information Foraging Theory provides a complementary account of how programmers navigate unfamiliar code bases \cite{storey2006theories, sillito2006questions}. From this perspective, programmers follow information scents, such as identifiers, file names, and call relationships, to locate code relevant to their current goals \cite{ko2006exploratory, lawrance2010programmers}. During maintenance tasks, programmers repeatedly locate focus points, inspect related code, and revise their understanding as they encounter new information.

These processes are especially challenging for students. Educational frameworks such as the Block Model describe comprehension across levels ranging from individual code elements to relationships among larger program structures \cite{schulte2008block}. However, students often struggle to navigate large code bases, identify relevant information, and distinguish task-relevant code from irrelevant implementation details \cite{shah2025needles}. Traditionally, successful program modification has therefore depended on programmers constructing a sufficiently accurate mental model of the relevant code \cite{shaft2006role}. GenAI coding assistants may alter this relationship by performing part of the implementation and reasoning work that programmers previously carried out themselves.

\subsection{Applications of GenAI in Code Comprehension}

GenAI has increasingly been used to support code comprehension in both educational and professional contexts. In computing education, researchers have asked students to explain code in natural language and then used an LLM to regenerate the original code from the explanation, treating the quality of the regenerated code as evidence of the completeness of the student's explanation
\cite{denny2024explaining, smith2024prompting}.
Other work has developed GenAI-based programming assistants such as CodeAid \cite{kazemitabaar2024codeaid}, embedded automatically generated explanations into educational materials \cite{macneil2023experiences}, and compared explanations generated by LLMs with those produced by students and experts \cite{leinonen2023comparing, sarsa2022automatic}.

Related tools have been designed to support professional programmers. GILT explains selected code without requiring users to construct detailed prompts \cite{nam2024using}, while Ivie provides lightweight explanations anchored to recently generated code \cite{yan2024ivie}. To address the limited context available in individual LLM interactions, Lei et al. proposed a multi-agent approach that generates explanations at the function, file, and project levels \cite{lei2025enhancing}. Brachman et al. further examined how programmers' goals and project contexts shape the explanations they need \cite{brachman2025towards}.

These studies demonstrate that GenAI can be deliberately designed to support program understanding. However, most examine systems that explicitly position the LLM as a comprehension aid. General-purpose assistants such as GitHub Copilot are primarily designed to generate and modify code. Evidence that an LLM can produce useful explanations therefore does not necessarily establish that programmers will request, process, or learn from those explanations while completing implementation tasks.

The closest work to ours is Shah et al.'s study of how students used GitHub Copilot while working with a large code base \cite{shah2025needles}. Their study characterized students' prompts and interactions with Copilot. Our work extends this research by directly assessing participants' comprehension after they complete brownfield programming tasks. This allows us to examine not only how programmers use Copilot, but also how their usage behaviors are associated with their resulting understanding of the codebase.

\subsection{GenAI's Impact on Productivity and Comprehension}

A growing body of research documents GenAI's effects on programming productivity. Controlled studies have found that GitHub Copilot can reduce task completion time and improve task success \cite{peng2023impact}. Recent work cautions, however, that these gains may be accompanied by concerns about the quality, reliability, and maintainability of AI-assisted code \cite{he2025speed}.

In computing education, the relationship between productivity and learning is particularly important. Prather et al. found that GenAI can provide substantial benefits to novice programmers while also creating new forms of dependence and widening differences between students who use the technology effectively and those who do not \cite{prather2024widening}. Studies of students' actual GenAI usage also reveal substantial variation in how learners prompt the tool, interpret its responses, and integrate generated code into their solutions \cite{rahe2025programming}.

This variation has motivated research on cognitive engagement with AI-generated code. Kazemitabaar et al. investigated interaction techniques that encourage learners to actively process generated solutions rather than immediately accept them \cite{kazemitabaar2025exploring}. Pu et al. similarly showed that AI assistance can support task progress while also interrupting or replacing programmers' own reasoning \cite{pu2025assistance}. Together, these studies suggest that the effects of GenAI may depend not only on whether the tool is available, but also on how programmers engage with its output.

Viewing an AI-generated response is not necessarily equivalent to understanding it. Programmers may inspect, revise, test, and connect generated code to their emerging mental models, or they may accept a plausible solution primarily because it compiles or passes automated tests. This distinction is particularly important in brownfield programming, where developers have traditionally had to understand how a modification fits the surrounding architecture and behavior. GenAI may help programmers determine what and where to modify while reducing the need to reason about how the resulting code operates.

Our study contributes empirical evidence to this productivity--comprehension tension. Rather than examining only whether Copilot improves task performance, we measure participants' code comprehension and analyze the programming behaviors associated with different comprehension outcomes. In particular, we investigate whether active behaviors such as writing, inspecting, and verifying code distinguish programmers with stronger comprehension from those who primarily consume AI-generated responses.

\section{Methods}
\label{sections-s3-methods}
\subsection{Overview and Research Design}
We conducted an IRB-approved within-subjects experimental study to investigate how GitHub Copilot influences graduate CS students' code comprehension and programming processes in brownfield programming tasks. Participants completed two isomorphic feature implementation tasks in an unfamiliar legacy codebase under two conditions: \textsc{No Copilot} (control) and \textsc{Copilot} (experimental).

\textbf{Key design decisions:} To capture participants' baseline programming behavior uninfluenced by GenAI, all participants completed the No Copilot condition first, followed by the Copilot condition. We counterbalanced task order (which feature was implemented in which condition) to mitigate task-specific effects. After completing tasks in each condition, participants completed a comprehension assessment designed to measure their understanding of the codebase they had just modified. 

This design extends prior work~\cite{shihab2025effects} by adding comprehension assessments and analyzing graduate-level students in brownfield contexts. Two minor procedural adaptations were made for this study's asynchronous online format: (1) instructions were delivered via Canvas rather than a setup lab, and (2) task timing was enforced using a custom Visual Studio Code extension that prevented access to subsequent tasks until the current task was submitted.

\subsection{Participants}
We recruited participants from a fully online, asynchronous graduate software engineering methods course with 44 enrolled students at a large research university in the western United States. Our final dataset comprised 15 graduate students (14 Master’s, 1 Ph.D.) who provided valid data; this includes three participants from the pilot study, as the protocol remained unchanged. The sample (12 male, 3 female) was high-achieving, with 13 participants holding a GPA of 3.7/4.0 or higher. Regarding GenAI usage frequency, five participants reported daily use, five weekly, three monthly, and two rarely (once or a few times). In terms of GenAI proficiency, five participants indicated high proficiency, eight medium, one low, and one reported no proficiency.

\subsection{Materials and Tasks}
\subsubsection{Legacy Codebase}
Participants worked within a 3,818-line subset of a sports analytics web application codebase, comprising 49\% JavaScript, 40\% HTML, and 11\% CSS. The application supported user authentication, profile management, and data tracking functionality, with all data stored in \texttt{localStorage}, eliminating the need for a back end.

\subsubsection{Feature Implementation Tasks}
Participants completed two isomorphic feature implementations, Add Distance and Add Picture, which required similar cognitive demands across three sequential subtasks: UI construction (adding form elements such as input fields, radio buttons, or image selectors), interactive behavior (implementing dynamic functionality like unit conversion or image preview), and data persistence (saving user data to localStorage).

Through a complexity analysis, we also confirmed comparable implementation difficulty: both solutions required similar lines of code (80 vs. 71), program statements (29 vs. 28), and control structures (3 vs. 3).

\subsubsection{Automated Test Suites}
Each feature included 13 automated tests distributed across the three subtasks (4 tests for Task 1, 4 for Task 2, 5 for Task 3). Tests verified UI rendering, behavioral correctness, and data persistence, providing a basis for measuring solution correctness.

\subsection{Comprehension Assessment Design}

To assess participants' understanding of the legacy codebase, we designed eight comprehension questions for each feature following established methodologies~\cite{sillito2006questions, stapleton2020human}. Each assessment included seven multiple-choice questions and one open-ended question covering five comprehension dimensions as shown in Table \ref{tab:combined_comprehension}.
\begin{table}[ht]
    \caption{Comprehension assessment design: Categories, focus, grading structure, and sample questions for the ``Add Picture'' task.}
    \label{tab:combined_comprehension}
    \centering
    \small

    \rowcolors{2}{gray!10}{white}

    \tagpdfsetup{table/header-rows={1}}

    \begin{tabularx}{\textwidth}{
        p{2cm}
        p{3.5cm}
        c
        c
        c
        X
    }
        \toprule
        \textbf{Category} &
        \textbf{Focus} &
        \textbf{Number} &
        \textbf{Points} &
        \textbf{Total} &
        \textbf{Sample Question} \\
        \midrule

        System Objective &
        High-level understanding of function goals &
        1 & 2 & 2 &
        What functionality is supported by the functions defined in
        \texttt{editProfile.js}? \\

        Implementation &
        Identification of specific files and logic flow &
        2 & 1 & 2 &
        To add a new error message for validating new incorrect patterns
        in a text field within the ``Account \& Profile'' form, which files
        would need to be modified to make it functional? \\

        Bug &
        Ability to identify errors in legacy context &
        2 & 2 & 4 &
        Imagine you are experiencing an issue where answer fields are not
        correctly resetting when exiting the ``Update Profile'' dialog.
        Which function would you look into to identify and fix the bug? \\

        Reverse Engineering &
        Predicting behavior if code were removed &
        2 & 2 & 4 &
        How would the application behavior change if we removed the code
        block from \texttt{editProfileForm.addEventListener()} in
        \texttt{editProfile.js}? \\

        Open-ended Question &
        Synthesis of function interactions &
        1 & 3 & 3 &
        Consider the three main functions in \texttt{editProfile.js}:
        \texttt{reset}, \texttt{populate}, and \texttt{update}.
        Briefly explain how these functions work together when users
        update their profiles. \\

        \midrule
        \multicolumn{4}{r}{\textbf{Total Score}} &
        \textbf{15} & \\
        \bottomrule
    \end{tabularx}
\end{table}

\textit{Scoring.} Multiple-choice questions were automatically scored, with point deductions for incorrect selections in multi-select items. Two authors independently scored open-ended responses using the rubric in Table~\ref{tab:grading-rubric}, achieving high inter-rater reliability (Krippendorff's $\alpha = 0.95$ for Control, $\alpha = 0.86$ for Experimental).

\begin{table}[htbp!]
\centering
\small
\caption{Grading rubric for comprehension questions}
\label{tab:grading-rubric}
\tagpdfsetup{table/header-rows={1}}
\rowcolors{2}{gray!10}{white}

\begin{tabular}{
  l
  p{0.15\columnwidth}
  p{0.60\columnwidth}
}\toprule
\textbf{Points} & \textbf{Descriptor} & \textbf{Description} \\
\midrule
3 & Exemplary & The answer correctly addresses all parts of the question. It demonstrates a thorough and clear understanding of the code base. \\
2 & Proficient & The answer correctly addresses the main part of the question, but is missing a minor detail or lacks some clarity. The core concept is understood, but the explanation is incomplete. \\
1 & Developing & The answer shows some relevant knowledge, but is fundamentally incomplete or contains a significant error. For instance, the student may have mentioned a function or concept but failed to explain it correctly. \\ 
0 & Incorrect & The answer is off-topic, completely incorrect, or left blank. \\
\bottomrule
\end{tabular}
\end{table}

\subsection{Procedure}
The 110-minute study sessions began with a 15-minute setup where participants signed an informed consent form, reviewed instructions emphasizing that Copilot (not them) was being evaluated, and completed a tutorial on the codebase and Copilot functionality. Participants then moved on to a 30-minute \textbf{No-Copilot} condition to implement the first feature without Copilot assistance, followed by a 30-minute \textbf{Copilot} condition for the second feature with the tool enabled, both concluding with a 10-minute comprehension assessment. The session wrapped up with a 15-minute post-survey regarding the differences in understanding the codebase and writing code between the two conditions.

\textit{Task Allocation.}  Each task had a 30-minute implementation limit (excluding time spent reading instructions). Timing was paused between subtasks to allow participants to read the next instructions without time pressure.

\textit{Counterbalancing.} While all participants completed No Copilot first (to capture baseline behavior), we counterbalanced which feature was implemented in which condition.

\subsection{Data Collection and Analysis}

\textit{Quantitative Analysis. }We captured task completion time in seconds for each subtask, the number of test cases passed, and comprehension scores. Quantitative analysis also included behavioral coding of programming activities (see Section 3.7) to categorize participants' actions rigorously.

\textit{Qualitative Analysis. }We performed a thematic analysis of the exit interview transcripts to identify user sentiments. We also gathered and examined responses to open-ended comprehension questions to provide a deeper context for the study's findings.

\textit{Statistical Analysis. }Given our small sample size (\textit{n}=15) and within-subjects design, we used non-parametric tests (Wilcoxon signed-rank tests for paired comparisons, Spearman correlations for associations), with Benjamini-Yekutieli corrections for multiple comparisons where appropriate.

\subsection{Behavioral Coding Scheme}
\label{sec:behavioralcoding}

\begin{table*}[!ht]
\caption{Coding Scheme for Programming Activities}
\label{tab:annotation-schema}
\tagpdfsetup{table/header-rows={1}}
\rowcolors{2}{white}{gray!15} 
\begin{tabular}{p{3cm}p{4cm}p{7cm}}
\toprule
\textbf{Primary Category} & \textbf{Secondary Category} & \textbf{Description} \\
\midrule
\rowcolor{white} 
\multirow{1}{*}{View Task} & -- & Searching or viewing task materials and instructions \\

\rowcolor{gray!15} 
\multirow{1}{*}{View Code} & -- & Searching or viewing code in editor or file explorer \\

\rowcolor{white} 
\multirow{1}{*}{View Web} & -- & Searching or viewing web materials for reference \\

\rowcolor{gray!15} 
\multirow{1}{*}{View App} & -- & Viewing the web application in browser \\

\rowcolor{white} 
\multirow{1}{*}{View Dev Tools} & -- & Viewing browser developer tools \\

\rowcolor{gray!15} 
\multirow{1}{*}{View Response} & -- & Viewing Copilot chat responses or autocomplete suggestions \\

\rowcolor{white} 
\multirow{-1}[0]{*}{Write Code} & Paste Copilot & Copy/paste code from Copilot chat \\
\rowcolor{white} & Accept Copilot Suggestion & Accept Copilot autocomplete suggestion (via tab) \\
\rowcolor{white} & Insert using Copilot& Insert code using inline Copilot chat interface (Copilot Edits) \\
\rowcolor{white} & Paste Web & Copy/paste code from web sources \\
\rowcolor{white} & Enter/Edit & Manually enter or edit code \\
\rowcolor{white} & Paste Self & Copy/paste from current code project \\
\rowcolor{white} & Modify Copilot Suggestion & Modify or delete recently accepted Copilot suggestion \\

\rowcolor{gray!15} 
\multirow{-1}[0]{*}{Test CLI} & Run Test & Running tests from command line \\
\rowcolor{gray!15} & View Test & Viewing test results \\

\rowcolor{white} 
\multirow{-1}[0]{*}{Write Prompt } & Paste Task & Copy/paste from task materials into prompt \\
\rowcolor{white} & Paste Previous Copilot Output & Copy/paste from previous Copilot response \\
\rowcolor{white} & Paste Error & Copy/paste from error messages \\
\rowcolor{white} & Enter Prompt & Manually enter/edit prompt \\
\rowcolor{white} & Enter Prompt as Comment & Enter prompt as code comment \\

\rowcolor{gray!15} 
\multirow{1}{*}{Idle} & -- & No active interaction occurring \\
\bottomrule
\end{tabular}
\end{table*}

To systematically analyze how programming processes differed with and without Copilot, we used an activity coding scheme from prior work~\cite{shihab2025effects} for iterative video analysis. The scheme, shown in Table \ref{tab:annotation-schema},  captures both primary activities (e.g., View Code (VC), Write Code (WC), Test Code (TC) and secondary categories providing granular detail about Copilot interactions (e.g., Accept Copilot Suggestion, Paste Copilot Response).
\vspace{2mm}

\textit{Reliability. }
Three authors independently coded a random 20\% sample of the participant videos, achieving a Krippendorff's $\alpha = 0.828$, which indicates strong agreement. The authors then divided the coding of the remaining videos. For the survey analysis, we manually coded the responses and engaged in iterative discussions among the authors until thematic saturation was achieved.

\section{Results}
\label{sections-s4-results}
\subsection{Performance Context }
\label{sec:replication}
 To provide context for our comprehension analysis, we first confirm that our study replicated the performance improvements found in prior work~\cite{shihab2025effects}. Participants completed Task 1 significantly faster with Copilot ($M=632.33s$) than without Copilot ($M=1281.5s$), representing a 50.7\% reduction in task completion time ($W=3.0,p<0.0001,r=0.836$). Solution correctness also increased by 71.4\%, with participants passing an average of 7.2 tests compared to 4.2 in the control condition ($W=6.5,p=0.004,r=0.772$). Replicating previous research \cite{shihab2025effects}, these findings establish that GitHub Copilot significantly enhanced both speed and correctness in our graduate student sample. We now turn to our primary research questions examining code comprehension.

\subsection{RQ1: Impact of GenAI on Code Comprehension}
\subsubsection{Overall Comprehension Levels}

Participants scored an average of 7.86 out of 15 points ($SD=2.82$, 52\% correct) without Copilot and 7.27 ($SD=3.25$, 48\% correct) with Copilot. A Wilcoxon signed-rank test showed this difference was not statistically significant ($W=50.5$, $p=0.59$, $r=0.14$). Analysis of individual results revealed that 53.3\% of participants scored lower with Copilot while 46.7\% improved; these changes were not statistically significant.

\subsubsection{Comprehension by Question Category}

Category-level analysis revealed no significant differences between conditions across System Objective, Implementation, Bug, Reverse Engineering, or Open-ended questions (all $p>0.05$, Wilcoxon signed-rank tests). While Implementation (66.67\% vs 60.00\%) and Reverse Engineering (56.67\% vs 50.00\%) showed slight declines with Copilot, these trends did not reach statistical significance. This uniform pattern across cognitive levels reinforces that Copilot's impact on comprehension was not limited to specific question types, but instead reflected a systematic lack of comprehension improvement.

\subsubsection{Participant Perceptions}

These quantitative findings align with participant perceptions: 10 of 15 participants reported that Copilot did not help them understand the application codebase. Several patterns emerged in their explanations:

\textit{Tool for generation, not comprehension.} Multiple participants explicitly stated they used Copilot for code generation rather than understanding. P20 explained: \textit{"I didn't really look to Copilot to explain the codebase, rather just help search through it."} 

\textit{Shallow explanations.} Participants who sought explanations found them insufficient. P31 noted: \textit{``Copilot did not explain much and focused mostly on the specific code block it had to fix to meet the requirements.''} P17 described feeling disconnected from the code: \textit{``It seems like Copilot likes providing entire functions without discussing what changes were made, which made me feel out of the loop.''}

\textit{Understanding requires engagement over time.} Some participants recognized that genuine comprehension requires hands-on experience. P43 observed: \textit{``I think codebase understanding comes after working with it for a bit.''} Most strikingly, P32 articulated the trade-off: \textit{``Copilot's tools helped me to write code more easily, and without having to think or understand what other parts of the codebase my code was interacting with. This didn't help me understand the codebase as a whole.''}

\textit{Bypassing comprehension for speed.} P36, who achieved the highest task performance (13/13 tests passed) but low comprehension (5.00/13), exemplified this pattern: \textit{``Copilot made it very easy, reduced it to copy and pasting the first solution it gave, which was the correct one"}, yet reported Copilot helped understanding only \textit{``just a little bit, not much.''}

\begin{tcolorbox}[colback=gray!10, colframe=black!70, 
                  width=\textwidth, arc=3mm, auto outer arc,
                  boxrule=0.5pt, drop shadow, fonttitle=\bfseries, 
                  title=Summary of RQ1 results]
GenAI assistance did not improve code comprehension in brownfield programming tasks, despite significant productivity gains. This finding contrasts with studies showing GenAI facilitates comprehension in greenfield programming contexts \cite{yan2024ivie}.
\end{tcolorbox}

\subsection{RQ2: Comprehension-Performance Correlation}
\subsubsection{Overall Correlation}

To investigate the relationship between comprehension and performance, we conducted a Spearman’s rank correlation analysis ($N=15$). The analysis revealed no statistically significant correlation ($\rho = -0.29, p = 0.29$), indicating that participants who completed more task requirements did not necessarily achieve higher comprehension scores.

This result reveals a \textit{comprehension-performance decoupling}: while participants demonstrated significant gains in task performance with Copilot (Section~\ref{sec:replication}), their comprehension did not improve commensurately. Thus, the traditional positive relationship between comprehension and performance observed in prior work~\cite{peitek2022correlates, shaft2006role} appears not to hold in GenAI-assisted coding.

Qualitative data strongly supports this comprehension-performance decoupling. High performers in the \textsc{No Copilot} condition explicitly described cognitive engagement with the codebase. P11, who passed 7 tests, noted that it was \textit{``difficult to figure out everything. Have to go through files properly... I have web development experience, so I understand how things work in the code.'`'} Similarly, P20 stated that \textit{``It took a while to find some specific parts of the codebase since some of the files are so large.''}

In contrast, high performers in the \textsc{Copilot} condition described minimal cognitive engagement. P36 reported: \textit{"Copilot made it very easy, reduced it to copying and pasting the first solution it gave, which was the correct one."} P26, who passed 8 tests but scored 4.50/15, explicitly acknowledged: \textit{"Was a lot faster in completing tasks but wasn't fully engaged with the code and didn't fully comprehend all the code I was putting in the files."} This qualitative evidence suggests that Copilot enabled high performance while bypassing code-level comprehension.

\subsubsection{Correlation by Category}



To explore whether performance improvement was associated differently with specific forms of comprehension, we calculated Spearman correlations between participants' change in task performance and their change in each comprehension category. For both measures, change was calculated as the Copilot-condition score minus the No-Copilot-condition score. Because we examined five comprehension categories, we applied the Benjamini--Yekutieli correction across the five correlations.

As shown in Table~\ref{tab:correlation}, Reverse Engineering produced the largest correlation estimate ($\rho=-0.570$, unadjusted $p=0.026$). Participants who obtained larger performance gains with Copilot tended to show smaller gains, or larger declines, in their ability to predict the consequences of changing or removing existing code. However, this association did not remain statistically significant after correction ($p_{\mathrm{BY}}=0.297$).

Implementation comprehension showed a non-signficant positive trend ($\rho=0.496$, unadjusted $p=0.060$), suggesting that participants with larger performance gains may have been more likely to retain knowledge about what needed to change and where the relevant code was located. However, with correction, this association landed even further from significance ($p_{\mathrm{BY}}=0.343$). The correlations for System Objective, Bug, and Open-ended comprehension were weaker and did not approach statistical significance.

\begin{table}[ht!]
\centering
\caption{Spearman correlations between task performance improvement and comprehension gain. The BY-adjusted $p$-values account for the five category-level comparisons.}
\label{tab:correlation}
\tagpdfsetup{table/header-rows={1}}
\small
\renewcommand{\arraystretch}{1.25}
\setlength{\tabcolsep}{4pt}
\rowcolors{2}{gray!10}{white}

\begin{tabularx}{\columnwidth}{
    >{\raggedright\arraybackslash}X
    r
    r
    r
}
\toprule
\rowcolor{white}
\textbf{Category}
& \multicolumn{1}{c}{\textbf{$\rho$}}
& \multicolumn{1}{c}{\textbf{Raw $p$}}
& \multicolumn{1}{c}{\textbf{BY-adjusted $p$}} \\
\midrule
System Objective
& -0.221 & 0.429 & 1.000 \\

Implementation
& 0.496 & 0.060 & 0.343 \\

Bug
& -0.071 & 0.801 & 1.000 \\

Reverse Engineering
& -0.570 & 0.026 & 0.297 \\

Open-ended
& 0.160 & 0.569 & 1.000 \\
\bottomrule
\end{tabularx}
\end{table}




Although these category-level relationships were not statistically significant, participants' post-survey responses illustrate how high task performance could coexist with low comprehension. P17, who passed 11 tests, described copying the task requirements into Copilot and receiving the necessary changes, but stated, \textit{``I just copy and pasted the requirements into the prompt and it gave me the necessary changes... I don't gain the understanding.''}. Similarly, P15 passed nine tests but scored 0 out of 4 on the Reverse Engineering questions, explaining: \textit{``It made it pretty easy to pick up and use my knowledge of the larger picture to get things done while the AI picked up on the smaller tasks.''}

These examples show that, in some cases, participants could achieve relatively high task performance without demonstrating strong knowledge of the codebase. Together with the correlation estimates, these observations motivate the exploratory hypothesis that Copilot may support task completion through knowledge of the desired behavior and reliance on generated implementation details. A larger study is needed to determine whether this pattern is reliable and whether it results from offloading reasoning, differences in prior ability, characteristics of the study tasks, or other factors.

\begin{tcolorbox}[
    colback=gray!10,
    colframe=black!70,
    width=\textwidth,
    arc=3mm,
    auto outer arc,
    boxrule=0.5pt,
    drop shadow,
    fonttitle=\bfseries,
    title=Summary of RQ2 Results
]
Copilot significantly improved task performance, but these improvements were not accompanied by measurable gains in overall code comprehension. None of the category-level correlations remained statistically significant after the Benjamini--Yekutieli correction. 

\end{tcolorbox}

\subsection{RQ3: Behavioral Predictors of Comprehension}

To explore behavioral patterns associated with code comprehension when using Copilot, we analyzed activity sequences from the 15 participants who completed Task 2 with Copilot assistance. Using a median split on final comprehension scores (median = 7.0/15), we compared high-comprehension participants ($n=8$, $M=10.12$, range: 7.00--11.98) with low-comprehension participants ($n=7$, $M=4.45$, range: 2.00--6.66). This 5.67-point difference (Cohen's $d=1.75$) represents a clear split between lower and higher participants' level of code comprehension.

\subsubsection{Verification Loop: The Strongest Predictor}

\begin{figure*}[t]
  \centering
  \includegraphics[width=\textwidth, alt={Two state-transition networks compare higher- and
lower-comprehension participants. Nodes represent programming
activities, with larger nodes indicating more session time and
thicker arrows indicating more frequent transitions. The
higher-comprehension group shows a prominent bidirectional
verification loop between Write Code and View Code. The
lower-comprehension group shows less activity in this verification
loop and greater concentration around Write Prompt and View Response.}]{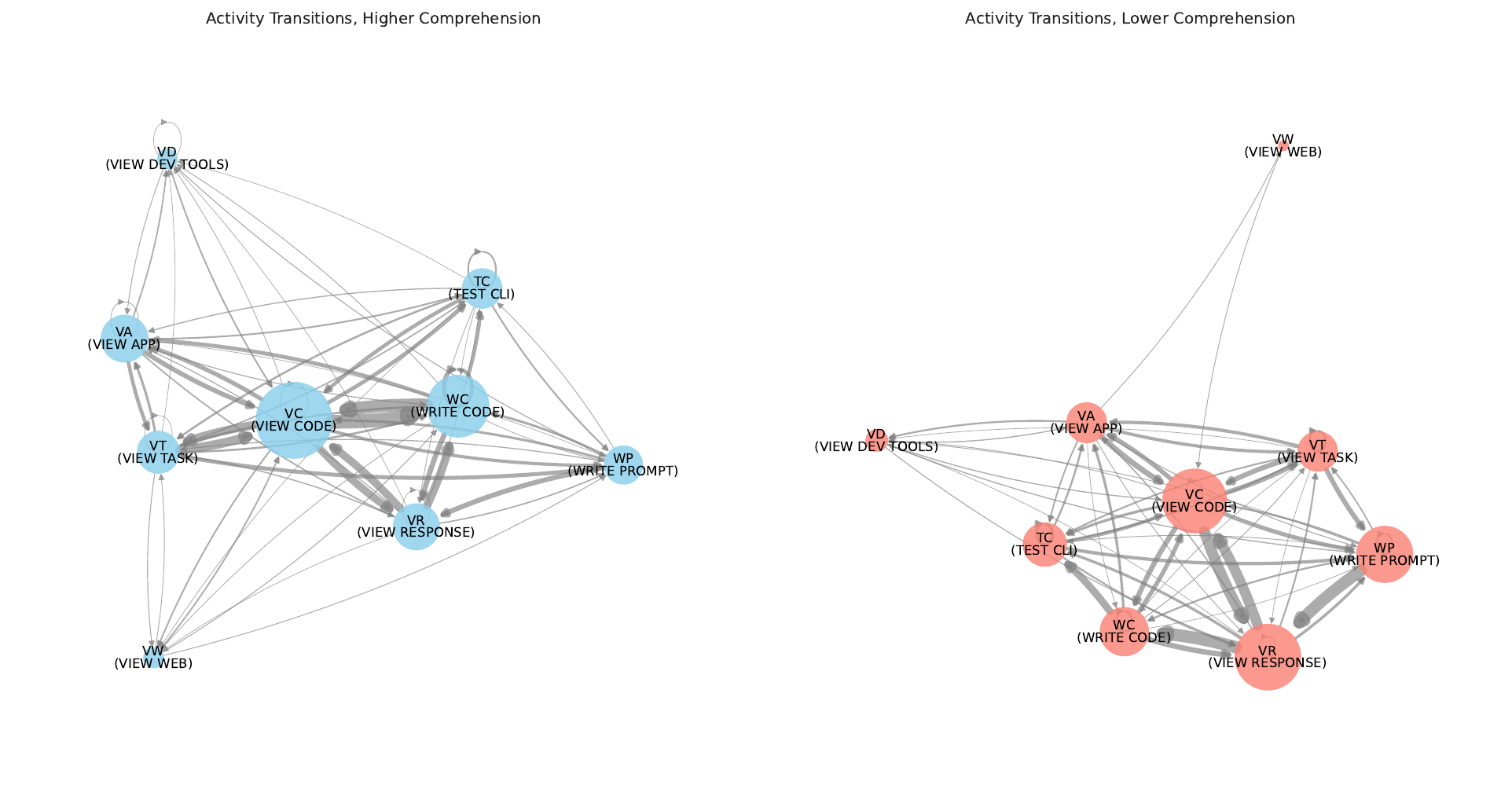}
  \caption{Probabilistic state-transition networks of programming activity in the
  Copilot condition, for higher-comprehension (left, $n=8$) and
  lower-comprehension (right, $n=7$) participants. Nodes are the primary activity
  codes, sized by the mean proportion of session time spent in each; arrows are
  directed transitions, with thickness proportional to mean transition frequency
  and self-loops indicating repeated coding of the same activity. Matrices are
  averaged across participants, so each participant contributes equally
  regardless of session length. Higher-comprehension participants show a dense
  \texttt{WC}$\leftrightarrow$\texttt{VC} verification loop, whereas
  lower-comprehension participants concentrate around
  \texttt{WP}$\rightarrow$\texttt{VR}.}
  \Description{Two node-link diagrams comparing activity transitions between
  higher- and lower-comprehension participants using GitHub Copilot. The
  higher-comprehension network shows thick bidirectional arrows between Write
  Code and View Code. The lower-comprehension network shows its heaviest traffic
  between Write Prompt and View Response, with View Code and View Response as the
  largest nodes.}
  \label{fig:transitions}
\end{figure*}

Figure \ref{fig:transitions} presents probabilistic state-transition networks depicting the programming activities of lower and higher comprehending participants.  As these figures suggest, higher-comprehension participants demonstrated significantly higher probabilities for both directions of the verification loop: viewing code after writing ($P(\text{WC→VC})=0.409$ vs $0.200$, $p=0.002$, $r=0.96$) and writing code after viewing ($P(\text{VC→WC})=0.329$ vs $0.156$, $p=0.032$, $r=0.68$). In practical terms, higher-comprehension participants performed the WC→VC transition 4.7 times more frequently than their lower-comprehension peers (10.8 vs 2.3 occurrences, $p=0.001$).

This pattern suggests that higher-comprehension participants engaged in active self-regulated learning: they did not simply accept Copilot's suggestions but instead repeatedly checked their work, creating a tight feedback loop between code generation and code inspection.

\subsubsection{Task-Driven Coding vs. Copilot-Driven Coding}

Higher-comprehension participants were significantly more likely to engage in \textit{task-driven coding}, in which they transitioned directly from viewing task requirements to writing code ($P(\text{VT→WC})=0.074$ vs $0.007$, $p=0.020$, $r=0.68$). This pattern indicates they maintained agency over the problem-solving process, starting with specification rather than immediately consulting Copilot.

In contrast, although lower-comprehension participants showed no significant preference for Copilot-initiated workflows with respect to transition probabilities, their time allocation revealed a notable dependency pattern: they spent nearly three times more time viewing Copilot responses (VR: 24.0\% vs 8.2\%, $p=0.004$, $r=0.88$). This suggests a \textit{passive consumption} pattern: Lower-comprehension participants spent their cognitive effort reading AI-generated output rather than actively writing or verifying code themselves.

\subsubsection{Active vs. Passive Programming Behaviors}

\begin{figure*}
    \centering
    \includegraphics[width=\textwidth, alt={Three bar charts compare higher- and lower-comprehension
participants. Higher-comprehension participants have a verification
loop probability of 0.738 compared with 0.357 for lower-comprehension
participants. They spend 20.6 percent of session time writing code
compared with 10.7 percent for the lower-comprehension group.
In contrast, they spend 8.2 percent of session time viewing Copilot
responses compared with 24.0 percent for the lower-comprehension
group. Error bars show standard deviations.}]{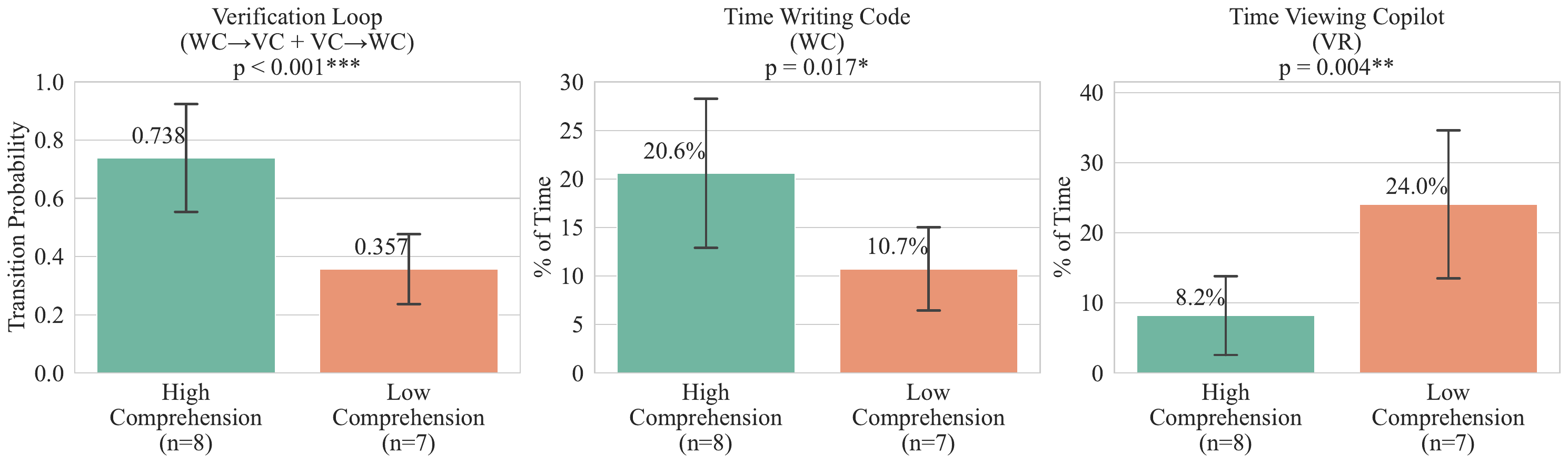}
    \caption{Key behavioral differences between high (n=8) and low (n=7) comprehension participants when using GitHub Copilot. High-comprehension participants exhibited higher verification loop engagement ($p<0.001$), spent more time writing code ($p=0.017$), and spent less time viewing Copilot responses ($p=0.004$). Error bars represent standard deviation.}
    \label{fig:key-behavioral}
\end{figure*}

Time allocation analysis revealed fundamental differences in how the two groups spent their programming sessions (Figure~\ref{fig:key-behavioral}). Higher-comprehension participants dedicated significantly more time to writing code (20.6\% vs 10.7\%, $p=0.017$, $r=0.76$) and conducted external research when needed (2.1\% vs 0.0\%, $p=0.011$, $r=0.71$). Overall, higher-comprehension participants exhibited 69\% more activity transitions (119 vs 71), indicating more dynamic engagement with the codebase.

Lower-comprehension participants, conversely, allocated their time primarily to more passive activities: viewing Copilot responses (24.0\% of time) and minimal code verification. This pattern aligns with a ``delegation'' workflow where understanding is offloaded to the GenAI tool rather than actively constructed through exploration and coding.

\subsubsection{Behavioral Profiles: Active Integrators vs. Passive Delegators}

\begin{figure}
    \centering
    \includegraphics[width=0.5\columnwidth, alt={Radar chart comparing higher- and lower-comprehension
participants across five programming behaviors: verifying code after
writing, writing code after viewing it, task-driven coding, external
research, and reviewing code after prompting. The
higher-comprehension group has higher values across all five
behaviors, with the largest differences in verifying code after
writing and writing code after viewing it. The lower-comprehension
profile remains closer to the center of the chart across the five
dimensions.}]{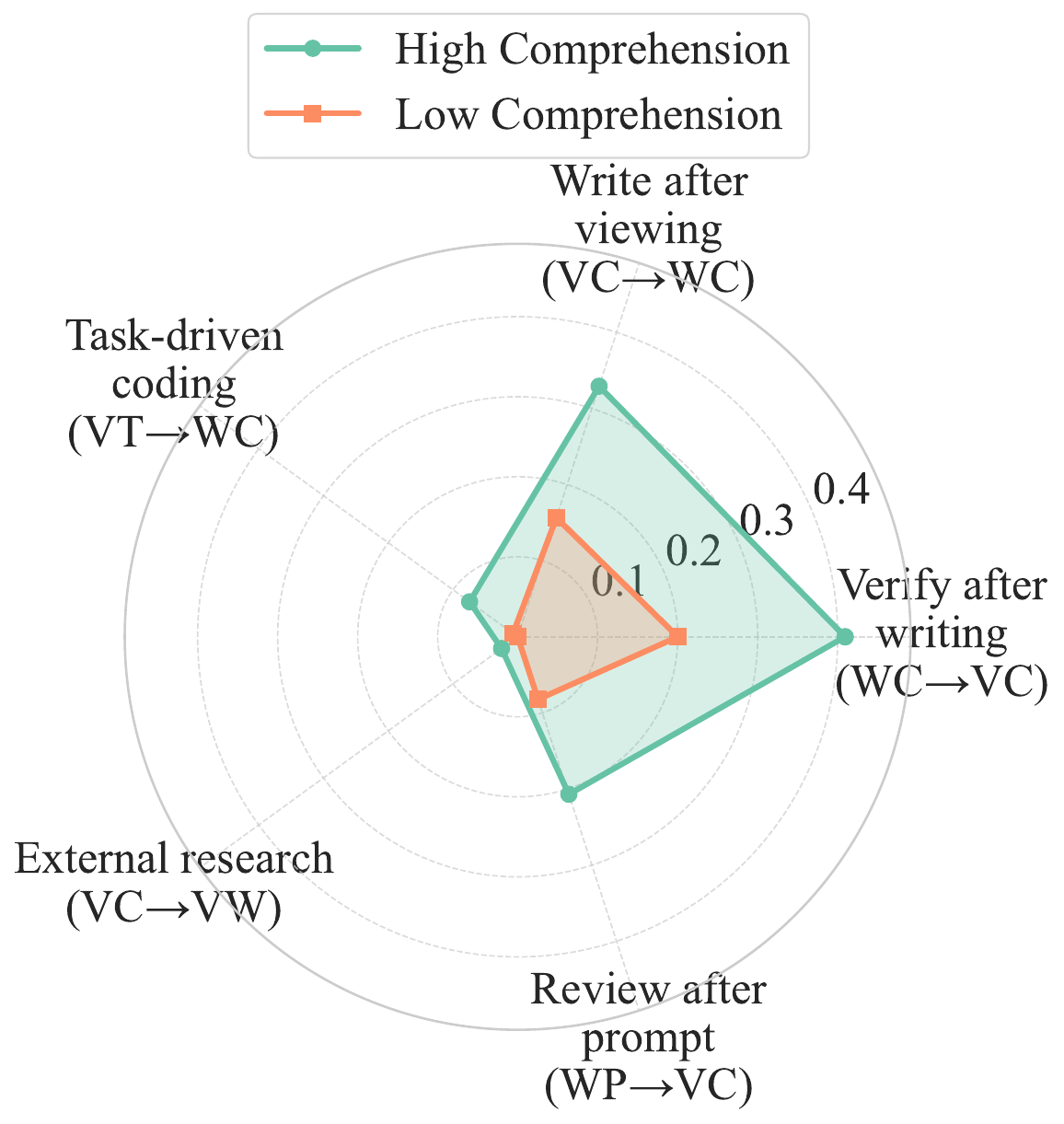}
    \caption{Radar chart of the behavioral profiles of higher vs. lower comprehension participants across five key programming activities. Higher-comprehension participants (green) demonstrate active integration through verification and task-driven coding, whereas lower-comprehension participants (orange) show less verification and task engagement.}
    \label{fig:behavioral-radar}
\end{figure}

Figure~\ref{fig:behavioral-radar} visualizes the distinct behavioral profiles that emerged. \textbf{Active Integrators} (higher comprehension) demonstrated:
\begin{itemize}\setlength{\itemsep}{0pt}\setlength{\parsep}{0pt}
    \item Frequent verification loops (2× more likely)
    \item Task-driven coding workflows (10× more likely)
    \item External research when needed
    \item Higher overall activity (69\% more transitions)
    \item Double the time spent writing code
\end{itemize}
\vspace{2mm}
In contrast, \textbf{Passive Delegators} (lower comprehension) showed:
\begin{itemize}\setlength{\itemsep}{0pt}\setlength{\parsep}{0pt}
    \item Minimal code verification
    \item Heavy reliance on viewing Copilot responses (3× time allocation)
    \item No external research
    \item Lower overall activity
    \item Less hands-on coding time
\end{itemize}

Qualitative data corroborate these patterns. Lower-comprehension participants explicitly described passive acceptance of GenAI output. P36 (score: 5.00/15) exemplified this approach: \textit{"Copilot made it very easy, reduced it to copying and pasting the first solution it gave, which was the correct one.''} P26 (score: 4.50/15) acknowledged the trade-off: \textit{''Was a lot faster in completing tasks but wasn't fully engaged with the code and didn't fully comprehend all the code I was putting in the files.''} This passive delegation pattern contrasts sharply with the verification behaviors observed in higher-comprehension participants' activity sequences.

\begin{tcolorbox}[colback=gray!10, colframe=black!70, 
                  width=\textwidth, arc=3mm, auto outer arc,
                  boxrule=0.5pt, drop shadow, fonttitle=\bfseries, 
                  title=Summary of RQ3 results]
The \textbf{verification loop} emerged as the strongest behavioral predictor of comprehension with Copilot ($p<0.001$, $r=0.96$). Participants who actively verified code achieved significantly higher comprehension than those who passively consumed AI-generated responses.
\end{tcolorbox}

\section{Discussion}
\label{sections-s5-discussion}
\subsection{RQ1: GenAI Does Not Automatically Improve Code Comprehension}

Our analysis found no significant difference in participants' overall code
comprehension between the No-Copilot (52\%) and Copilot (48\%) conditions. This
pattern was consistent across the different categories of comprehension
questions. Although Copilot substantially improved task completion speed
and correctness, these productivity gains were not accompanied by a
corresponding improvement in participants' understanding of the codebase. The result is consistent with evidence from recent studies \cite{balepur2026paired, alakmeh2026grasping}.

This result suggests that in GenAI-assisted software development, improved productivity does not automatically
translate into improved comprehension. In conventional brownfield
programming, developers generally need to inspect existing code, trace
relationships, and construct a mental model sufficiently accurate to implement
a modification. These activities require programmers to move iteratively
among different parts of a system and across multiple levels of
abstraction~\cite{levy2019understanding, von1997program}.
By contrast, Copilot can generate a plausible modification from a task
description and the local code context. This reduces the amount of
exploration and implementation reasoning that the programmer must perform on their own. Participants can therefore complete more of the task
without necessarily engaging in the same comprehension process that
modification would require.

Importantly, this finding should not be interpreted as evidence that GenAI
is inherently incapable of supporting code comprehension. Participants
used a GenAI coding assistant for implementation tasks; they were not explicitly instructed to use Copilot as a
comprehension aid. Hence, their immediate objective was to implement the
requested functionality within the time budget. Under these conditions,
participants had a strong incentive to use Copilot to reduce
implementation effort rather than to request explanations or
systematically explore the codebase. Indeed, most participants reported
that they treated Copilot primarily as a code-generation tool rather than
as a tool for learning about the system.

The result may also reflect a mismatch between Copilot's local code generation ability and the multi-layered nature of program comprehension. Copilot
can quickly produce a function or localized code change. Understanding a
brownfield system, however, requires programmers to connect that change to
related files, event sequences, data flows, and component relationships.
Three participants explicitly noted that Copilot focused on the specific code
needed to satisfy the requirements without explaining how that code fit
into the broader system. Unless programmers actively reconstruct this
broader context, localized generation may accelerate implementation
without supporting the mental-model construction needed for system
understanding.

The fixed condition order provides an additional perspective on this
result. Participants completed the Copilot condition after already working
with the same application in the No-Copilot condition. They therefore
entered the Copilot task with greater familiarity with the project
structure, interface, and development environment. This familiarity could
have raised their comprehension scores. However, no improvement was observed.
Although the fixed order prevents us from fully separating tool effects
from familiarity, the confound works against the effect we might have
expected. 

These findings raise a broader concern for computing education. If, with GenAI tools,
students can produce more correct code without developing a
commensurate understanding of that code, code correctness
becomes an unreliable indicator of learning. Automated tests establish
that an implementation satisfies selected behavioral requirements. They do
not establish that the student can explain the implementation, anticipate
its side effects, or maintain the code when the requirements change. The emergence of GenAI programming tools therefore makes it important to evaluate programming
performance and code comprehension as related but distinct outcomes.

\subsection{RQ2: Performance Gains Were Not Reliably Coupled with Comprehension Gains}

Our RQ2 results show that the performance improvements participants obtained from Copilot were not reliably accompanied by stronger code comprehension. Copilot significantly increased the number of tests participants passed, but overall comprehension did not improve correspondingly, and task performance and overall comprehension in the Copilot condition were not significantly correlated. This result does not establish that performance and comprehension are fully independent. Instead, it shows that, in our sample, successful task completion was not a reliable indicator that a participant understood the code they had modified. Related studies report a similar pattern. Participants using a coding agent to build a web application outperformed a comparison group on the initial task while scoring lower on comprehension questions about their own code~\cite{balepur2026paired}. Similarly, developers modifying an unfamiliar full-stack codebase demonstrated comparable comprehension across AI and no-AI conditions despite differences in task outcomes~\cite{alakmeh2026grasping}.

At the category level, none of the associations between GenAI-assisted performance gains and comprehension scores were statistically significant after the Benjamini–Yekutieli correction. We therefore found no statistically reliable evidence that performance gains were associated differently with particular forms of comprehension.

Several alternative explanations remain plausible. Because the analysis used change scores, participants' initial scores and the amount of room they had to improve may have influenced the estimates. Participants who performed well without Copilot had less room to improve, whereas participants who initially struggled could obtain larger performance gains. If initial ability also influenced how participants used Copilot or how their comprehension changed, it could have contributed to the observed negative estimate. Other factors also remain uncontrolled. Participants differed in prior programming ability, how much they relied on Copilot, and the quality of the suggestions they received. In addition, the bounded nature of the tasks may have reduced the amount of understanding required for successful completion. The observed correlations alone therefore cannot show that participants delegated reasoning about the code to Copilot or that such delegation reduced their comprehension.

A further possibility is that successful GenAI use depended more on specifying \textit{what} the program should do than on independently determining \textit{where} and \textit{how} to implement it. Participants may have achieved strong performance by describing the required behavior clearly, asking Copilot to locate the relevant code, and steering the assistant when its response was incomplete. Their advantage would then come from communicating the task with little ambiguity and judging whether the generated response satisfied the requirements, rather than from knowing the modification point in advance. This workflow would represent an alternative pathway to successful program modification. In conventional brownfield programming, developers generally construct a mental model of the relevant code and use that model to guide a change~\cite{roehm2012professional, shaft2006role}. With GenAI, programmers may instead specify the desired behavior, ask the assistant to locate and modify the relevant code, and evaluate the result through automated tests. Our comprehension scores cannot determine whether participants supplied the modification locations, learned them from Copilot, or relied on Copilot to locate them.

Participants' comments illustrate how high performance and limited understanding of the modified code could coexist. P17 reported copying the requirements into Copilot and receiving the necessary changes, but stated, ``I don't gain the understanding.'' P15 passed nine tests while receiving no points on the Reverse Engineering questions, explaining that Copilot allowed them to use their knowledge of the larger picture while the AI handled the smaller tasks. Both abilities assessed in our study matter in brownfield work. Identifying what code needs to change and where the change belongs remains an important software engineering ability, particularly in large systems~\cite{sillito2006questions}, and it supports immediate task completion. Understanding how the relevant code works supports debugging, explanation, maintenance, and adaptation. The limitations of not fully understanding an implementation may not appear during a well-specified task when the generated solution passes the provided tests, but recent evidence suggests that they may become visible later. Participants who used a coding agent showed no advantage when later asked to extend their own code without assistance~\cite{balepur2026paired}. Programmers who learned an unfamiliar library with AI assistance scored lower on a subsequent unaided assessment, with the largest difference appearing on debugging items~\cite{shen2026ai}. At the project level, velocity gains following the adoption of an AI coding tool were confined to the first two months, while increases in code complexity persisted~\cite{he2025speed}. A programmer may therefore complete the immediate task successfully while remaining less prepared for subsequent maintenance work.

The category-level results should be interpreted as exploratory. The small sample, limited score ranges within individual categories, and lack of statistical significance after multiple-comparison correction prevent us from concluding that Copilot redistributes comprehension or causes a decline in Reverse Engineering ability. Future analyses should examine how much of the final code came from Copilot suggestions, whether participants supplied or requested modification locations in their prompts, and how they inspected, verified, and revised generated code. Our data also include the proportion of session time each participant spent viewing Copilot responses. This measure would allow us to explore whether the relationship between performance improvement and changes in Reverse Engineering comprehension becomes weaker after accounting for time spent viewing Copilot responses. Such a result would be consistent with the explanation that reliance on Copilot contributed to the observed pattern, although it would not establish causality. Larger studies should also use delayed comprehension assessments and transfer tasks that require participants to debug or extend the generated implementation. Such evidence would help distinguish reasoning delegation from differences in initial ability, effective task specification, and characteristics of the study tasks.

\subsection{RQ3: Verification Distinguishes Active Integration from Passive Delegation}

\begin{figure}[htbp]
    \centering
    \includegraphics[width=0.5\columnwidth, alt={Flowchart of a proposed Copilot-assisted programming workflow.
The process begins by viewing task requirements and writing code based
on those requirements. The programmer may then write a prompt and
evaluate or integrate the GenAI-generated code. If more understanding
is needed, the programmer consults external resources such as web
pages or documentation and returns to the implementation process.
The programmer iterates between writing and evaluating code until the
implementation is complete, then tests the code. Failed tests return
the programmer to code revision, while passing tests lead to task
completion.}]{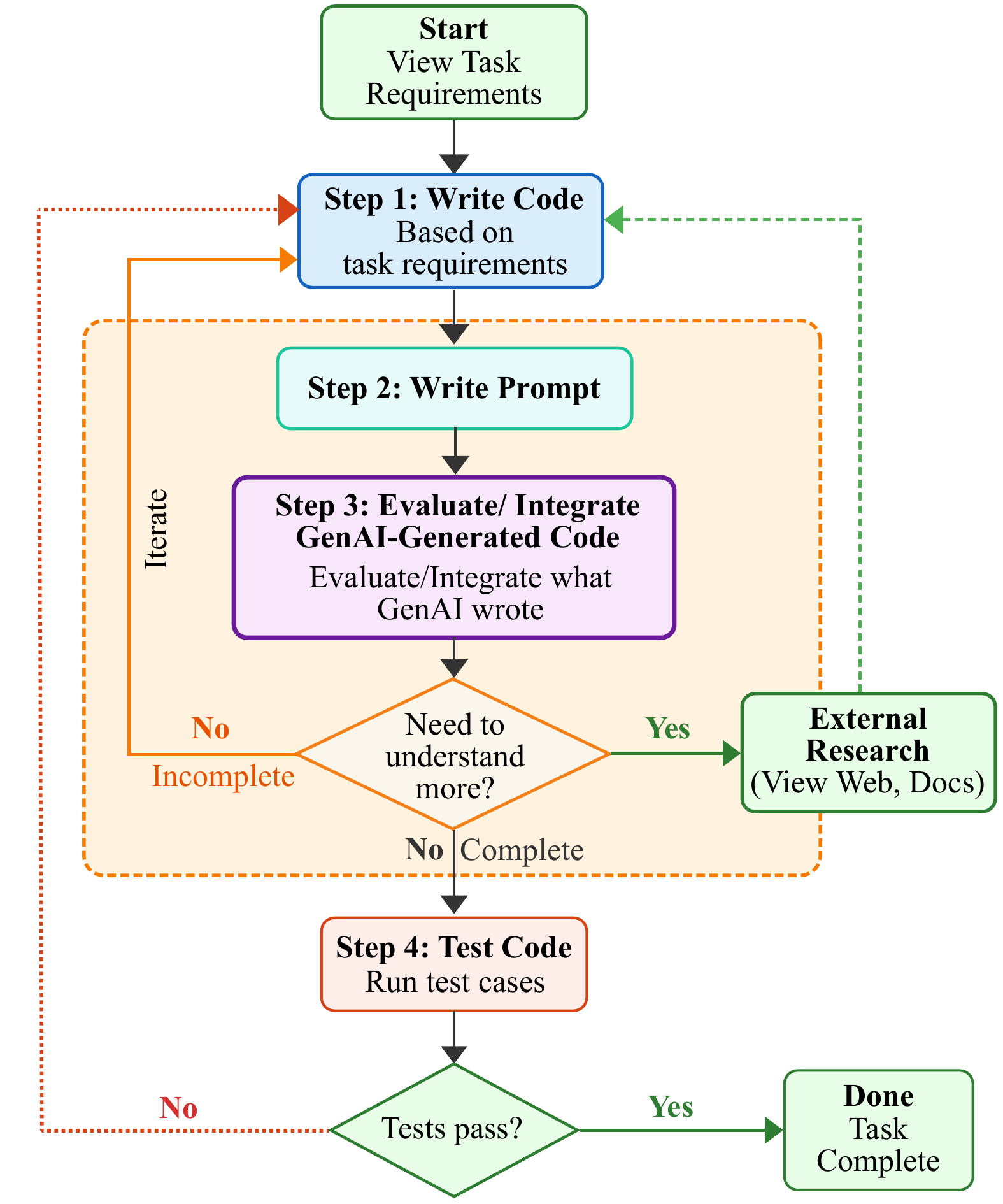}
    \caption{Evidence-informed workflow for supporting comprehension during Copilot-assisted brownfield programming. The workflow synthesizes behaviors associated with higher comprehension in our study. The highlighted verification loop was the behavioral pattern most strongly associated with comprehension (\(p < 0.001\), \(r = 0.96\)). Because these relationships are associative, the workflow represents a hypothesis for future evaluation rather than a validated intervention.}
    \label{fig:recommended-workflow}
\end{figure}

Our behavioral analysis provides insight into why similar access to Copilot produced different comprehension outcomes. Participants with higher comprehension engaged much more frequently in verification loops, repeatedly moving between writing or integrating code and inspecting the resulting implementation. The combined verification-loop probability was more than twice as high for the high-comprehension group, and they performed the write-then-view transition 4.7 times more frequently than participants with lower comprehension.

Verification may support comprehension because it requires programmers to reconcile generated code with their current understanding of the system. When participants returned to inspect code after writing or integrating it, they had an opportunity to compare the generated implementation with existing patterns, trace how it interacted with surrounding functions, identify inconsistencies, and revise their assumptions. Similarly, moving from code inspection back to writing required participants to act on what they had learned. The educational value of the verification loop may therefore lie not in the physical transition between activities itself, but in the cognitive reconciliation that the transition represents.

This interpretation is consistent with prior research on cognitive engagement with AI-generated code. Activities such as prediction, explanation, modification, and reflection require learners to process generated solutions rather than immediately accepting them
\cite{kazemitabaar2025exploring}. Proactive assistance can accelerate task progress, but it may also interrupt or replace the programmer's own reasoning
\cite{pu2025assistance}. Our findings provide complementary behavioral evidence: participants who repeatedly inspected and revised code demonstrated stronger comprehension than participants whose interaction was dominated by viewing Copilot responses.

Notably, lower-comprehension participants spent nearly three times as much of their session viewing Copilot responses. This result indicates that exposure to AI-generated information should not automatically be interpreted as cognitive engagement. A participant may spend substantial time reading a response without tracing its logic, comparing it with the existing implementation, or evaluating the assumptions on which it depends. In contrast, editing generated code, revisiting related files, and alternating between inspection and implementation require the programmer to incorporate the response into the task context.

The behavioral results also distinguish comprehension-oriented verification from reliance on automated tests alone. Running tests provides evidence that selected behavioral requirements have been satisfied, but it does not necessarily require the programmer to understand why the implementation succeeds. Tests can function as an external acceptance signal: if the tests pass, the generated code may be treated as correct. Comprehension-oriented verification additionally involves examining implementation logic, considering dependencies and side effects, and connecting the modification to the surrounding system. This distinction helps explain why participants could achieve high test performance while demonstrating limited understanding of the generated code.

Higher-comprehension participants also spent more time writing code and were more likely to transition directly from viewing task requirements to writing. These behaviors suggest that they retained greater agency over the implementation process. Rather than immediately asking Copilot to solve the entire task, they first attempted to interpret or act on the requirements themselves. When they used Copilot, its output became part of an ongoing problem-solving process rather than a replacement for that process. Their occasional use of external resources further suggests that they recognized gaps in their knowledge and sought additional information instead of relying exclusively on the generated response.

By contrast, lower-comprehension participants' behavior was more consistent with delegation. They spent less time writing code, conducted little external research, and devoted a larger share of their sessions to viewing Copilot output. Several participants explicitly described copying or accepting the first plausible solution. This workflow can be effective for immediate task completion, especially when the generated solution is correct, but it provides fewer opportunities for programmers to construct and revise their own mental models.

We therefore describe the two patterns as \textit{active integration} and \textit{passive delegation}. Active integration occurs when programmers treat GenAI output as material to inspect, modify, test, and connect to their understanding of the system. Passive delegation occurs when programmers transfer substantial implementation reasoning to the GenAI and evaluate its output primarily through plausibility or test outcomes. These terms describe modes of interaction rather than fixed categories of programmers. The same participant may actively integrate one suggestion and passively delegate another depending on prior knowledge, task difficulty, time pressure, and the apparent quality of the generated output.

The behavioral findings are associative rather than causal. Verification loops showed the strongest relationship with comprehension among the behaviors we examined, but the study does not establish that verification caused participants to understand the code better. Participants with stronger prior understanding or greater programming ability may also have been more likely to verify generated code. Nevertheless, the strength and consistency of the observed differences identify verification as an important candidate mechanism for future experimental study.

Although these associative findings do not establish a causal effect of verification, they motivate an evidence-informed workflow for preserving comprehension during GenAI-assisted programming. Figure~\ref{fig:recommended-workflow} translates the behaviors associated with higher comprehension into a provisional workflow: programmers begin by interpreting the task requirements, inspect and integrate GenAI-generated changes through repeated verification loops, test the resulting implementation, and consult external resources when additional understanding is needed. This workflow should be interpreted as a hypothesis for future evaluation rather than as a validated intervention.

\section{Implication}
\label{sections-s6-implication}
\subsection{Implications for Programming Education}

Our findings suggest that programming performance alone may not provide sufficient evidence of code comprehension in GenAI-assisted programming. In our study, Copilot significantly increased the number of tests participants passed, but overall comprehension did not improve correspondingly, and task performance and comprehension in the Copilot condition were not significantly correlated. This result does not establish that performance and comprehension are independent. Instead, it shows that, in our sample, successful task completion was not a reliable indicator that a participant understood the code they had modified. In conventional programming, successful modification generally requires programmers to locate relevant code, reason about its behavior, and construct a mental model that guides implementation. With GenAI, students may be able to communicate the desired functionality, integrate generated code, and validate the result through automated tests without developing an equally detailed understanding of the implementation. Although our study does not establish that all participants followed this pathway, it shows that successful task completion did not reliably indicate comprehension.

Prior work characterizes program comprehension as involving several activities, including locating relevant code, expanding understanding around that code, identifying relationships among program elements, and connecting those relationships at the system level~\cite{sillito2006questions}. Our category-level correlation estimates showed different patterns for Implementation and Reverse Engineering comprehension. However, neither association remained statistically significant after correction for multiple comparisons. The results therefore do not establish that GenAI redistributes students' comprehension or causes them to offload particular forms of reasoning. Instead, they motivate the possibility that performance gains may not be accompanied equally by the ability to identify what and where to change and the ability to explain how the relevant code works and predict the effects of a modification.

This possibility does not mean that either ability is no longer important. Although GenAI can generate implementation details, students remain responsible for determining whether those details are correct, appropriate for the surrounding codebase, and maintainable over time. A student who can identify where a modification belongs but cannot explain how the generated code works may struggle to debug unexpected behavior, adapt the code to a new requirement, or assess side effects that are not covered by the available tests. Programming education should therefore continue to develop students' abilities to locate relevant code, explain how it works, and reason about the consequences of changing it, rather than replacing these abilities with prompt construction alone.

At the same time, curricula should increasingly emphasize students' ability to reason about software systems and relationships among their components. When GenAI is used to produce local code changes, programmers may spend more of their effort identifying relevant components, communicating requirements, evaluating proposed changes, and integrating those changes into an existing application. These abilities have been identified as increasingly important for software developers working in the age of AI~\cite{kam2025professional}. Educators can support their development through brownfield programming activities that require students to navigate unfamiliar codebases, trace dependencies, analyze the effects of proposed changes, and explain how a modification fits within the broader system.

However, teaching these abilities should not be reduced to teaching students how to write more effective prompts. Prior work has emphasized the importance of context-aware prompting for software development~\cite{liang2025prompts, nam2025prompting}. Prompting skills can help students communicate requirements and provide relevant codebase information, but a well-written prompt does not necessarily demonstrate that the student understands the resulting implementation. A sophisticated prompt may still delegate most of the reasoning and implementation work to the assistant.

Our behavioral findings suggest that programming education should also emphasize what students do after receiving GenAI output. The verification loop---repeatedly moving between writing or integrating code and inspecting the resulting implementation---showed the strongest association with comprehension among the behaviors we examined. Participants with higher comprehension did not merely receive generated code; they reviewed it, revised it, and repeatedly returned to inspect the surrounding implementation. These activities may help students connect generated solutions to their understanding of the codebase and preserve opportunities for learning. Because these behavioral relationships were associative rather than causal, further studies are needed to determine whether encouraging verification directly improves comprehension.

We therefore argue that programming education should develop students' \textit{AI code stewardship}: the ability to critically inspect, explain, test, modify, and take responsibility for AI-generated code within an existing software system. AI code stewardship extends beyond obtaining useful output. It requires students to determine why the output works, whether it fits the existing application, what assumptions it makes, and what consequences it may have elsewhere in the system.

Educators could support AI code stewardship by requiring students to predict which files and functions should change before consulting GenAI, explain generated code before accepting it, identify dependencies affected by a modification, or revise part of an AI-generated solution manually. Students could also compare a generated solution with existing implementation patterns, identify unsupported assumptions, or debug intentionally flawed AI-generated code. These activities would position GenAI output as material for reasoning rather than as a final answer.

\subsection{Implications for Programming Assessment}

The finding that performance gains were not reliably accompanied by comprehension gains also has direct implications for assessment. In GenAI-assisted programming environments, artifact correctness and student comprehension should be treated as related but distinct outcomes. Automated tests can determine whether submitted code satisfies selected requirements, but they cannot determine whether the student understands why the implementation works. A student may pass all available tests by successfully integrating a generated solution while remaining unable to explain its control flow, data dependencies, or interactions with other parts of the system.

Consequently, assessments based only on task completion, execution correctness, or the number of passed tests may overestimate learning. These measures remain valuable for evaluating whether an artifact functions correctly, but they should be supplemented with evidence of the student's understanding. This is particularly important in brownfield tasks, where successful long-term maintenance requires developers to understand existing code and reason about interactions beyond the immediate feature requirements.

Assessment designs could combine several forms of evidence. Artifact-level evidence can measure whether the implementation is correct and satisfies the task requirements. Comprehension evidence~\cite{lehtinen2021let} can assess whether students can explain generated code, predict its behavior, identify relevant dependencies, or reason about what would happen if particular code were changed or removed. Process evidence~\cite{zhang2026editrail} can document how students evaluated and revised GenAI output, including rejected suggestions, manual modifications, and verification strategies. Finally, transfer evidence~\cite{balepur2026paired} can assess whether students can debug or extend a related part of the system after completing the original task.

For example, students could submit an annotated diff explaining the purpose of each generated change and how it interacts with the existing code. Instructors could use short oral code defenses in which students explain a selected function, identify possible side effects, or respond to a hypothetical change in requirements. Assessments could also include Reverse Engineering questions that ask students to predict how the application would behave if a generated code block were changed or removed. These forms of assessment directly examine whether students can explain how generated code works and predict how changes to it would affect program behavior---abilities that were not reliably indicated by task performance in our study.

Educators may also ask students to document their use of GenAI, but prompt transcripts should be interpreted carefully. The presence of detailed or iterative prompts does not by itself demonstrate comprehension. A student may provide extensive context while still delegating reasoning about the code and its implementation to the tool. Based on our behavioral findings, evidence of active evaluation---such as inspecting, revising, explaining, and testing generated code---may be more educationally meaningful than prompt complexity alone.

Assessment incentives may also influence how students use GenAI. When grades reward only rapid completion and functional correctness, students have little reason to spend time developing a detailed understanding once a generated solution passes the tests. In contrast, when assessments explicitly reward explanation, reasoning about the effects of changes, and verification, students may be more likely to inspect and evaluate generated code. Assessment design can therefore shape not only what learning is measured but also how students interact with GenAI while completing a task.

\subsection{Implications for GenAI Tool Design}

The current generation of GenAI coding assistants primarily optimizes code generation and task completion rather than comprehension support~\cite{rodrigues2025redesigning}. Participants in our study frequently received localized code solutions without broader explanations of how the suggested changes related to the surrounding system. This design can reduce implementation effort, but it may also allow users to integrate code without fully understanding how the generated change works, how it interacts with other parts of the system, or what consequences it may have for later maintenance.

Future GenAI programming assistants should therefore support not only code production but also active inspection, integration, and verification. One possible direction is a \textit{Comprehension Mode} \cite{gao2026understanding} that supplements generated code with information about why the modification is needed, how it interacts with existing components, and what other parts of the system may be affected. Rather than explaining only the generated statements, such a mode could connect the suggestion to relevant functions, files, data flows, and relationships among components.

Tools could also encourage users to make predictions before receiving a complete solution. For example, an assistant might ask which files or functions the programmer expects to modify or request a brief description of the relevant data flow. Making a prediction could activate the programmer's existing understanding and make differences between the programmer's expectations and the generated solution more visible. After generating the solution, the assistant could highlight these differences and explain why additional components were affected.

Another design direction is progressive disclosure. Instead of immediately displaying an entire generated function, an assistant could present the solution in meaningful segments and allow users to inspect the purpose and dependencies of each segment before continuing. Prior work on cognitive engagement suggests that decomposing generated responses and prompting learners to actively process them may preserve opportunities for reasoning~\cite{kazemitabaar2025exploring, pu2025assistance}. This approach may be especially useful when a large generated response would otherwise encourage users to copy or accept the solution as a whole.

GenAI assistants could further support the verification loop identified in our behavioral analysis. After generating code, the tool might prompt the user to inspect the modified region, identify an assumption made by the solution, or describe how the change should be tested. It could show relevant callers, dependencies, or state changes and ask the programmer to consider possible side effects. When users accept a large generated change without opening or editing it, the assistant could remind them to review the modification and its relationship to the surrounding code.

Tools should also distinguish behavioral validation from comprehension-oriented verification. Running an automated test confirms selected outputs, but it does not necessarily help the programmer understand the implementation. A comprehension-oriented assistant could explain which execution paths a test exercises, identify behaviors not covered by the tests, or ask the programmer to predict a test result before execution. Such features could turn testing from a simple acceptance signal into an opportunity to reason about how and why the code produces its behavior.

These proposed features should be treated as design hypotheses rather than validated solutions. Additional research is needed to determine whether prediction prompts, progressive disclosure, dependency explanations, and verification support improve comprehension without imposing excessive interaction costs or eliminating GenAI's productivity benefits. Nevertheless, our findings suggest that future coding assistants should evaluate success not only in terms of how quickly users obtain working code but also in terms of whether users remain actively involved in understanding and taking responsibility for the generated implementation.

\section{Threats to Validity}
\label{sections-s7-threats-to-validity}
\subsection{Internal Threats to Validity}

Several factors threaten internal validity. First, participants completed and recorded the study without live monitoring, so some may not have followed all instructions or remained fully focused. The VS Code extension enforced task order and time limits, while the screen recordings allowed us to examine participants' activities retrospectively. However, these measures could not completely ensure compliance or sustained attention. Second, thinking aloud and recording the screen may have changed how participants worked or interacted with Copilot. Third, all participants completed the No-Copilot condition before the Copilot condition. Therefore, the effect of Copilot cannot be fully separated from order effects, including increased familiarity with the codebase, practice, and fatigue. We counterbalanced which feature appeared in each condition to reduce task-specific effects, but this did not remove the condition-order confound. Increased familiarity could have supported later performance or comprehension, while fatigue could have affected them in the opposite direction; therefore, the direction of this threat is uncertain. Finally, participants were instructed to complete the programming tasks but were not explicitly asked to understand the code while working. Their behavior may therefore reflect a task-completion goal rather than a comprehension goal. Different results may occur when programmers are told that they will need to explain, debug, or extend the resulting code.

\subsection{Construct Validity}

Our comprehension assessment included eight questions covering five categories. Although the questions assessed several aspects of code comprehension, a short assessment may not capture the full depth of participants' understanding. In particular, individual categories contained only one or two questions, resulting in limited score ranges and reducing the sensitivity of the category-level analysis. The Implementation questions assessed whether participants knew what and where to modify after completing the task, but they could not determine whether participants located the relevant code independently or learned its location through Copilot. Similarly, the Reverse Engineering questions assessed participants' ability to explain code behavior and predict the effects of removing code, but they did not capture every form of reasoning involved in understanding an unfamiliar codebase. Future studies should use more questions per category, delayed assessments, and transfer tasks requiring participants to debug or extend the generated implementation. Task performance was measured by the number of automated tests passed. This measure captures whether selected behavioral requirements were satisfied, but it does not capture code quality, maintainability, or correctness beyond the tested cases. Passing more tests should therefore be interpreted as improved performance on the study tasks rather than as evidence of overall programming success.

\subsection{Statistical Conclusion Validity}

The study included only 15 participants, limiting statistical power and the precision of the estimated relationships. A nonsignificant correlation should therefore not be interpreted as evidence that performance and comprehension are independent. The category-level analysis also correlated changes in task performance with changes in comprehension. Change scores can be influenced by baseline scores, ceiling or floor effects, regression to the mean, and measurement error. Because five category-level correlations were examined, we applied the Benjamini--Yekutieli correction. None remained statistically significant after correction; therefore, the different estimates for Implementation and Reverse Engineering comprehension should be treated as exploratory patterns rather than confirmed effects. The behavioral analyses also identify associations rather than causal relationships. For example, participants with higher comprehension engaged more frequently in verification loops, but the study does not establish that verification caused stronger comprehension. Participants with greater prior ability may have been both more likely to verify generated code and more likely to understand it.

\subsection{External Threats to Validity}

The findings may not generalize beyond graduate students from a single university. Participants' educational backgrounds, programming experience, and familiarity with GenAI may differ from those of undergraduate students, novice programmers, or professional developers. The study also used two feature-implementation tasks in a 3,818-line JavaScript web application. The explicit requirements, localized changes, automated tests, and short study duration may have allowed participants to complete tasks without developing the understanding required in larger or less well-specified projects. Conversely, larger systems may offer different opportunities for Copilot to support code navigation and explanation. Studies with industry-scale codebases, ambiguous requirements, incomplete tests, and longer maintenance activities are needed before the findings can be generalized to professional brownfield development. Finally, the study examined GitHub Copilot in a particular task setting. Other GenAI assistants, interaction modes, model versions, and levels of codebase access may produce different performance, comprehension, and behavioral outcomes.

\section{Conclusions}
\label{sections-s8-conclusions}
This study examined how GitHub Copilot influenced performance, comprehension, and programming behavior in brownfield tasks. Participants completed tasks faster and passed more tests with Copilot, but these gains were not accompanied by detectable improvements in code comprehension. Although the category-level estimates suggested different patterns for Implementation and Reverse Engineering comprehension, none remained significant after multiple-comparison correction. Participants with higher comprehension engaged more frequently in verification loops, repeatedly inspecting and revising code. These findings suggest that task success does not reliably demonstrate comprehension and that active engagement with generated code may matter. Programming education should therefore assess correctness and comprehension separately, teach students to inspect and explain AI-generated code, and encourage GenAI tools that support active verification.


\bibliographystyle{ACM-Reference-Format}
\bibliography{other}

\end{document}